\documentclass[twocolumn, tighten]{aastex62}

\usepackage{nicefrac}
\usepackage{microtype}
\usepackage{booktabs}
\usepackage{rotating}

\newcommand{\kms}{${\rm km\,s}^{-1}$}

\newcommand{\pks}{${\rm PKS}\,0405{-}123$}

\received{?}
\revised{?}
\accepted{?}
\submitjournal{ApJL}

\shorttitle{Galaxy and Quasar Fueling Caught in the Act}
\shortauthors{Johnson et al.}

\begin{document}

\title{Galaxy and Quasar Fueling Caught in the Act from the Intragroup to the Interstellar Medium}

\correspondingauthor{Sean D. Johnson}
\email{sdj@astro.princeton.edu}

\author[0000-0001-9487-8583]{Sean D. Johnson}
\altaffiliation{Hubble \& Carnegie-Princeton fellow}
\affil{Department of Astrophysical Sciences, 4 Ivy Lane, Princeton University, Princeton, NJ 08544, USA}
\affil{The Observatories of the Carnegie Institution for Science, 813 Santa Barbara Street, Pasadena, CA 91101, USA}

\author[0000-0001-8813-4182]{Hsiao-Wen Chen}
\affiliation{Department of Astronomy \& Astrophysics, The University of Chicago, 5640 S. Ellis Avenue, Chicago, IL 60637, USA}

\author[0000-0001-5892-6760]{Lorrie A. Straka}
\affil{Leiden Observatory, Leiden University, PO Box 9513, NL-2300 RA Leiden, the Netherlands}

\author[0000-0002-0668-5560]{Joop Schaye}
\affil{Leiden Observatory, Leiden University, PO Box 9513, NL-2300 RA Leiden, the Netherlands}

\author{Sebastiano Cantalupo}
\affil{Department of Physics, ETH Zurich, Wolfgang-Pauli-Strasse 27, 8093, Zurich, Switzerland}

\author[0000-0001-5020-9994]{Martin Wendt}
\affil{Leibniz-Institut f\"{u}r Astrophysik Potsdam (AIP), An der Sternwarte 16, 14482 Potsdam, Germany}
\affil{Institut f\"ur Physik und Astronomie, Universit\"at Potsdam, Karl-Liebknecht-Str. 24/25, 14476 Golm, Germany}

\author[0000-0001-5020-9994]{Sowgat Muzahid}
\affil{Leiden Observatory, Leiden University, PO Box 9513, NL-2300 RA Leiden, the Netherlands}

\author[0000-0003-0068-9920]{Nicolas Bouch\'{e}}
\affil{Univ Lyon, Univ Lyon1, Ens de Lyon, CNRS, Centre de Recherche Astrophysique de Lyon UMR5574, F-69230, Saint-Genis-Laval, France}

\author{Edmund Christian Herenz}
\affil{Department of Astronomy, Stockholm University, AlbaNova University Centre, 106 91 Stockholm, Sweden}

\author{Wolfram Kollatschny}
\affil{Institut f\"{u}r Astrophysik, Universit\"{a}t G\"{o}ttingen, Friedrich-Hund Platz 1, D-37077 G\"{o}ttingen, Germany}

\author[0000-0003-2083-5569]{John S. Mulchaey}
\affil{The Observatories of the Carnegie Institution for Science, 813 Santa Barbara Street, Pasadena, CA 91101, USA}

\author[0000-0002-8559-6565]{Raffaella A. Marino}
\affil{Department of Physics, ETH Zurich, Wolfgang-Pauli-Strasse 27, 8093, Zurich, Switzerland}

\author[0000-0003-0695-4414]{Michael V. Maseda}
\affil{Leiden Observatory, Leiden University, PO Box 9513, NL-2300 RA Leiden, the Netherlands}

\author{Lutz Wisotzki}
\affil{Leibniz-Institut f\"{u}r Astrophysik Potsdam (AIP), An der Sternwarte 16, 14482 Potsdam, Germany}



\begin{abstract}
We report the discovery of six spatially extended ($10{-}100$ kpc) line-emitting nebulae in the $z\,{\approx}\,0.57$ galaxy group hosting \pks,
one of the most luminous quasars at $z\,{<}\,1$. The discovery is enabled by the Multi Unit Spectroscopic Explorer (MUSE) and provides tantalizing evidence connecting large-scale gas streams with nuclear activity on scales of ${<}\,10$ proper kpc (pkpc). One of the nebulae exhibits a narrow, filamentary morphology extending over $50$ pkpc toward the quasar with narrow internal velocity dispersion ($50$ \kms) and is not associated with any detected galaxies, consistent with a cool intragroup medium (IGrM) filament. Two of the nebulae are  $10$ pkpc North and South of the quasar with tidal arm like morphologies. These two nebulae, along with a continuum emitting arm extending $60$ pkpc from the quasar are signatures of interactions which are expected to redistribute angular momentum in the host interstellar medium (ISM) to facilitate star formation and quasar fueling in the nucleus. The three remaining nebulae are among the largest and most luminous [O\,III] emitting ``blobs'' known ($1400{-}2400$ pkpc$^2$) and correspond both kinematically and morphologically with interacting galaxy pairs in the quasar host group, consistent with arising from stripped ISM rather than large-scale quasar outflows. The presence of these large- and small-scale nebulae in the vicinity of a luminous quasar bears significantly on the effect of large-scale environment on galaxy and black hole fueling, providing a natural explanation for the previously known correlation between quasar luminosity and cool circumgalactic medium (CGM).
\end{abstract}



\section{Introduction} \label{section:intro}

Galaxy$-$galaxy interactions represent one of the few cosmologically viable mechanisms for redistributing angular momentum in the ISM to fuel luminous quasars and nuclear star formation \citep[][and references therein]{Hopkins:2009}. In cosmological simulations of galaxy evolution, mergers play a significant role in fueling black hole growth at $z\,{<}\,1$ \citep[e.g.][]{McAlpine:2018}.
Despite these expectations and over fifty years of observations, the importance of interactions in fueling quasars is still debated with studies finding evidence both against \citep[e.g.][]{Villforth:2014} and in favor \citep[e.g.][]{Goulding:2018} of interactions as a major triggering mechanism.

Insights into quasar fueling can be gained through observations of gas in quasar host environments. Observations through H\,I $21$-cm emission are largely limited to the local Universe
while quasar activity peaked at $z\,{\approx}\,2$ \citep[e.g.][]{Schmidt:1995} leaving few available targets. More sensitive surveys using background absorption spectroscopy reveal the common presence of cool (${\approx}10^4$ K) circum-galactic medium (CGM) in quasar host halos \citep[][]{Bowen:2006, Hennawi:2006, Prochaska:2013, Farina:2014, Johnson:2015b} at projected distances of $d\,{\lesssim}\,300$ pkpc. This cool CGM exhibits extreme kinematics
and is strongly correlated with quasar luminosity, suggesting a physical connection between quasar activity and the CGM at $z\,{\approx}\,1$
\citep[for a study of the CGM of low-luminosity AGN, see][]{Berg:2018}.

The lack of morphological information in absorption-line surveys makes it difficult to differentiate between cool CGM often observed around massive galaxies \citep[e.g.][]{Chen:2018}, debris from interactions thought to fuel nuclear activity \citep[e.g.][]{Villar-Martin:2010}, and  outflows \citep[e.g.][]{Greene:2012}. Even when morphologies of extended nebulae around quasars are available from imaging \citep[e.g.][]{Stockton:1987, Sun:2017} or narrow-field Integral Field Spectrographs (IFS) \citep[e.g.][]{Fu:2009, Liu:2013, Husemann:2013}, discerning the origins of the nebulae can be difficult. Nevertheless, such emitting ``blobs'' are often attributed to outflows \citep[e.g.][]{Fu:2009, Schirmer:2016, Yuma:2017}.

New, wide-field IFSs such as MUSE \cite[][]{Bacon:2010} provide a powerful means of simultaneously surveying the galactic and gaseous environments of quasars allowing both sensitive searches for extended, ionized nebulae and joint studies of their morphologies and kinematics in the context of neighboring galaxies.
MUSE already enabled the discovery of extended nebulae around AGN in the field \citep[][]{Powell:2018}, in group or cluster environments \citep[][]{Poggianti:2017, Epinat:2018}, and around luminous quasars at $z\,{\approx}\,3$ \citep[e.g.][]{Borisova:2016}.

Here, we present the discovery of ionized nebulae on scales of $10{-}100$ pkpc in the environment of \pks, one of the most luminous quasars in the $z\,{<}\,1$ Universe\footnote{PKS\,0405$-$123 at $z\,{=}\,0.5731$ has a bolometric luminosity of $L_{\rm bol}\,{\approx}\,3 {\times}10^{47}\ {\rm erg\,s^{-1}}$ and a high inferred Eddington ratio of ${\sim}1$ \citep[][]{Punsly:2016}.}. Joint analyses of the nebular morphologies and kinematics indicate that they arise from cool filaments and interaction related debris rather than outflows. These observations provide novel insights into galaxy and quasar fueling from IGrM to ISM scales. 

This letter proceeds as follows: In Section \ref{section:observations} we describe the MUSE observations and analysis. In Section \ref{section:environment}, we present the galactic environment of \pks. In Section \ref{section:nebulae}, we present the discovery of multiple extended nebulae around the quasar and discuss their origins. In Section \ref{section:discussion} we consider the implications of our findings.

Throughout, we adopt a flat $\Lambda$ cosmology
with $\Omega_{\rm m}\,{=}\,0.3$, $\Omega_\Lambda\,{=}\,0.7$,
and $H_0\,{=}\,70\,{\rm km\,s^{-1}\,Mpc^{-1}}$.

\section{Observations and data}
\label{section:observations}

We obtained MUSE observations in the field of \pks\ as part of the MUSE Quasar-field Blind Emitter Survey (MUSE-QuBES), a guaranteed time observation program (GTO) on the Very Large Telescope (PI: J. Schaye, PID: 094.A-0131). The MUSE-QuBES motivations, survey strategy, and analysis will be detailed in Segers et al. and Straka et al., (in preparation). The data are briefly summarized here.

MUSE is an IFS with a $1'{\times}1'$ arcmin field-of-view (FoV), spectral coverage of $4750{-}9350$ \AA, and resolution of $R{=}2000{-}4000$ \cite[][]{Bacon:2010}. We acquired $9.75$ hours of MUSE integration for the field of \pks\ in October$-$November, 2014 under median full width at half maximum (FWHM) seeing of $0.7''$ and reduced the data using GTO reduction \citep[][]{Weilbacher:2014} and sky subtraction \citep[][]{Soto:2016} tools. We identified continuum sources in the field with Source Extractor \citep{Bertin:1996} using both a white-light image from the MUSE datacube and an image from the Advanced Camera (ACS) for Surveys
aboard {\it Hubble Space Telescope} (HST) with the F814W filter (PI: Mulchaey, PID: 13024). For each source, we extracted a $1$D spectrum using MPDAF \citep[][]{Piqueras:2017} and measured initial redshifts with MARZ \citep[][]{Hinton:2016}. In the process, we discovered multiple extended nebulae at redshifts similar to the quasar which contaminate some redshift measurements. Consequently, we re-extracted the galaxy spectra with $0.7''$ diameter apertures, masked strong emission lines, and measured the redshifts whenever possible based purely on stellar absorption by fitting SDSS galaxy eigenspectra \citep[][]{Bolton:2012}. The resulting galaxy redshift uncertainties are ${\approx}20$ \kms.

\begin{figure*}
	\includegraphics[width=\textwidth]{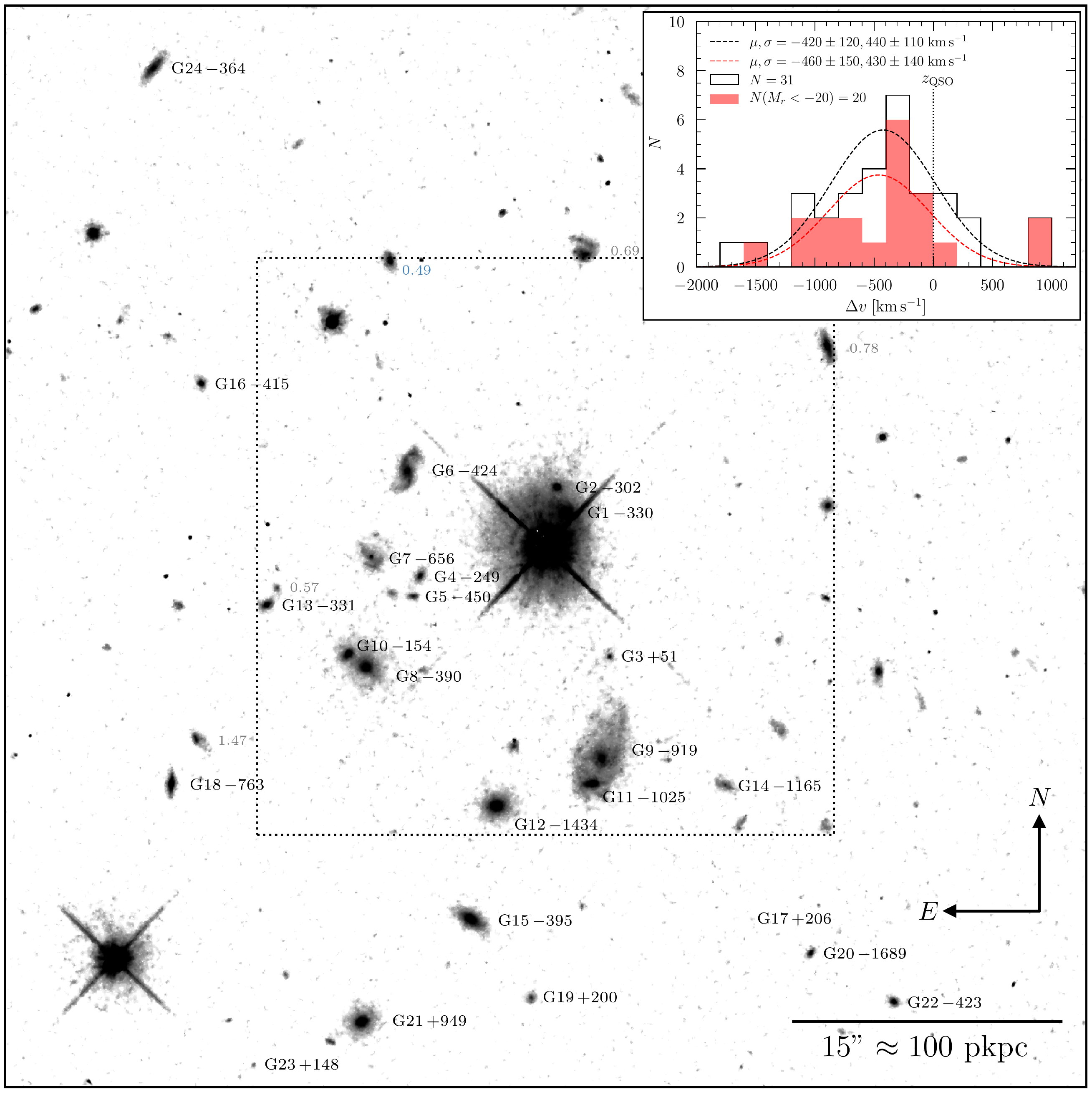}
	\caption{HST ACS/F814W image of the field of PKS\,0405$-$123.  Galaxies in the quasar host group are labelled by their ID and line-of-sight velocity from the quasar ($z\,{=}\,0.5731$) in \kms. Galaxies that are foreground (background) to the group are labelled in smaller font by their redshift from \protect \cite{Johnson:2013} in blue (grey). The image shows the $60''{\times}60''$ MUSE field-of-view and the dotted square marks the $30''{\times}30''$ region displayed in Figure \ref{figure:narrows}. The inset panel displays the line-of-sight velocity histogram of galaxies in the environment of PKS\,0405$-$123 with the zero corresponding to the quasar systemic redshift.}
	\label{figure:HST}
\end{figure*}

The brightness of \pks\ and broad wings of the MUSE point-spread function (PSF) result in contribution of light from the quasar to spaxels at ${\lesssim}8''$ from the quasar. However, \pks\ cannot be used to model the PSF because the host galaxy biases the model and stars in the field are not bright enough to measure the PSF wings.

To subtract the quasar light, we developed a technique that takes advantage of the spectral dimension provided by an IFS and the fact that galaxy and quasar spectra are distinct \citep[see also][]{Rupke:2017}. The primary challenge with this approach is the wavelength dependence of the PSF which disperses blue light further away from the quasar than red light, resulting in an
artificially flat (steep) quasar spectrum close to (far from) the quasar.  To account for this, we determined the two non-negative spectral components that can best model the quasar contribution to any spaxel as linear combinations by performing non-negative matrix factorization \citep[][]{Zhu:2016} on quasar dominated spaxels from a $1''{\times}1''$ aperture centered on the quasar. We then modeled each spaxel at ${<}8''$ from the quasar as a linear combination of the two quasar components and the first two SDSS galaxy eigenspectra shifted to $z\,{=}\,0.57$ with emission lines masked. We subtracted the quasar component of the best-fit model from each spaxel, effectively removing the quasar light contribution except at ${\lesssim}1''$ from the quasar.

\begin{table*}
\centering
\caption{Summary of galaxies in the field of PKS\,0405$-$123 at $z\,{\approx}\,z_{\rm QSO}$.}
\label{table:galaxies}
\begin{tabular}{cccccccrrrc}
\hline
ID    & R.A.        & Dec.      & $m_{\rm F814W}$ &  \multicolumn{1}{c}{$z$}     &   $u{-}g$   & \multicolumn{1}{c}{$M_{r}$}        & \multicolumn{1}{c}{$\Delta \theta$} &   \multicolumn{1}{c}{$d$} &  \multicolumn{1}{c}{$\Delta v$} &  redshift reference\\
        & (J2000)  & (J2000)  &  (AB)                     &                                            & (AB) & (AB) & \multicolumn{1}{c}{(arcsec)}           &  \multicolumn{1}{c}{(pkpc)} & \multicolumn{1}{c}{($\rm km\,s^{-1}$)} &  \\
\hline
\hline
host & 04:07:48.48 &${-}$12:11:36.0 & $-$  &$0.5731$ & $-$ & $-$ & 0.0 & 0.0 & 0 & this work \\
G1  & 04:07:48.40 & ${-}$12:11:34.2 & 21.3 & 0.5714 & $1.5$ & $-$21.3 & 2.2  & 14.3  & $-$330  & this work  \\
G2  & 04:07:48.43 & ${-}$12:11:32.8 & 21.7 & 0.5715 & $1.5$ & $-$20.9 & 3.3  & 21.9  & $-$302  & this work  \\
G3  & 04:07:48.24 & ${-}$12:11:42.2 & 23.6 & 0.5734 & $1.3$ & $-$19.0 & 7.1  & 46.5  & $+$51  & this work  \\
G4  & 04:07:48.95 & ${-}$12:11:37.7 & 22.8 & 0.5718 & $1.4$ & $-$19.8 & 7.1  & 46.8  & $-$249  & this work  \\
G5  & 04:07:48.98 & ${-}$12:11:38.8 & 23.8 & 0.5707 & $1.4$ & $-$18.8 & 7.9  & 51.4  & $-$450  & this work  \\
G6  & 04:07:49.00 & ${-}$12:11:31.8 & 21.1 & 0.5709 & $1.2$ & $-$21.3 & 8.8  & 57.3  & $-$424  & this work  \\
G7  & 04:07:49.13 & ${-}$12:11:36.7 & 21.9 & 0.5697 & $1.1$ & $-$20.5 & 9.7  & 63.2  & $-$656  & this work  \\
G8  & 04:07:49.16 & ${-}$12:11:42.7 & 20.7 & 0.5711 & $1.7$ & $-$21.7 & 12.0 & 78.6  & $-$390  & this work  \\
G9  & 04:07:48.26 & ${-}$12:11:47.8 & 20.0 & 0.5683 & $1.4$ & $-$22.3 & 12.2 & 79.9  & $-$919  & this work  \\
G10 & 04:07:49.22 & ${-}$12:11:42.0 & 21.5 & 0.5723 & $1.6$ & $-$20.9 & 12.5 & 81.7  & $-$154  & this work  \\
G11 & 04:07:48.30 & ${-}$12:11:49.2 & 20.9 & 0.5677 & $1.3$ & $-$21.4 & 13.4 & 87.7  & $-$1025 & this work  \\
G12 & 04:07:48.66 & ${-}$12:11:50.4 & 20.5 & 0.5656 & $1.7$ & $-$22.1 & 14.7 & 95.9  & $-$1434 & this work  \\
G13 & 04:07:49.53 & ${-}$12:11:39.3 & 22.3 & 0.5714 & $1.4$ & $-$20.3 & 15.8 & 103.4 & $-$331  & this work  \\
G14 & 04:07:47.80 & ${-}$12:11:49.3 & 22.8 & 0.5670 & $1.5$ & $-$19.5 & 16.6 & 108.4 & $-$1165 & this work  \\
G15 & 04:07:48.76 & ${-}$12:11:56.8 & 20.9 & 0.5710 & $1.8$ & $-$21.7 & 21.1 & 138.2 & $-$395  & this work  \\
G16 & 04:07:49.78 & ${-}$12:11:27.0 & 22.6 & 0.5709 & $1.6$ & $-$20.0 & 21.2 & 138.5 & $-$415  & this work  \\
G17 & 04:07:47.72 & ${-}$12:11:56.7 & 26.5 & 0.5742 & $0.8$ & $-$15.9 & 23.5 & 153.9 & $+$206  & this work  \\
G18 & 04:07:49.89 & ${-}$12:11:49.2 & 21.6 & 0.5691 & $1.7$ & $-$20.9 & 24.6 & 161.0 & $-$763  & this work  \\
G19 & 04:07:48.53 & ${-}$12:12:01.1 & 23.4 & 0.5742 & $1.5$ & $-$19.2 & 25.1 & 164.1 & $+$200  & this work  \\
G20 & 04:07:47.47 & ${-}$12:11:58.6 & 23.4 & 0.5642 & $1.6$ & $-$19.2 & 26.9 & 176.4 & $-$1689 & this work  \\
G21 & 04:07:49.17 & ${-}$12:12:02.4 & 20.7 & 0.5781 & $1.7$ & $-$22.0 & 28.3 & 185.1 & $+$949  & this work  \\
G22 & 04:07:47.16 & ${-}$12:12:01.3 & 22.8 & 0.5709 & $1.6$ & $-$19.8 & 31.8 & 208.3 & $-$423  & this work  \\
G23 & 04:07:49.58 & ${-}$12:12:04.8 & 25.5 & 0.5739 & $0.8$ & $-$16.9 & 33.0 & 216.2 & $+$148  & this work  \\
G24 & 04:07:49.96 & ${-}$12:11:09.5 & 21.8 & 0.5712 & $1.6$ & $-$20.8 & 34.3 & 224.8 & $-$364  & this work  \\
G25 & 04:07:49.43 & ${-}$12:12:10.8 & 21.0 & 0.5777 & $1.7$ & $-$21.6 & 37.5 & 245.4 & $+$877  & \cite{Ellingson:1994a}  \\
G26 & 04:07:48.76 & ${-}$12:12:18.6 & 21.4 & 0.5726 & $1.7$ & $-$21.1 & 42.8 & 279.9 & $-$95  & \cite{Johnson:2013}    \\
G27 & 04:07:46.63 & ${-}$12:12:09.8 & 22.0 & 0.5725 & $1.4$ & $-$20.6 & 43.3 & 283.3 & $-$114  & \cite{Johnson:2013}    \\
G28 & 04:07:45.99 & ${-}$12:10:59.8 & 20.1 & 0.5685 & $1.6$ & $-$22.3 & 51.5 & 336.9 & $-$877  & \cite{Chen:2009}       \\
G29 & 04:07:49.27 & ${-}$12:12:26.3 & 22.8 & 0.5692 & $1.7$ & $-$19.6 & 51.5 & 337.4 & $-$743  & \cite{Johnson:2013}    \\
G30 & 04:07:46.50 & ${-}$12:12:35.1 & 22.2 & 0.5675 & $0.7$ & $-$20.4 & 65.7 & 430.4 & $-$1067 & \cite{Johnson:2013}    \\
\hline
\end{tabular}
\end{table*}

\section{The galactic environment of \pks}
\label{section:environment}
We identified candidate members of the quasar host environment by selecting galaxies with line-of-sight velocities of $|\Delta v|\,{<}\,2000$ \kms\ from the quasar, $z_{\rm QSO}\,{=}\,0.5731{\pm}0.0003$ (measured from the [O\,II] line), including both our new MUSE catalog and galaxies outside the MUSE FoV from the ACS$+$F814W image with redshift measurements from the literature \cite[see][]{Johnson:2013}. We chose this velocity window to be approximately twice the velocity dispersion of the most massive galaxy clusters. We identified 31 (25) galaxies in the HST (MUSE) field within this velocity range including the quasar host.

For each galaxy, we report the right ascension (R.A.), declination (Dec.), observed ACS$+$F814W magnitude ($m_{\rm F814W}$), redshift ($z$), rest-frame $u\,{-}\,g$ color measured in matched isophotal apertures, rest-frame absolute $r$-band magnitude ($M_r$), and the projected angular ($\Delta \theta$), physical ($d$) and line-of-sight velocity ($\Delta v\,{=}\,v\,{-}\,v_{\rm QSO}$) differences from the quasar in Table \ref{table:galaxies}. Figure \ref{figure:HST} displays the ACS$+$F814W image of the field with group members labelled.

The quasar host environment includes four (twenty) galaxies of $M_r\,{<}\,{-}22$ (${<}\,{-}20$), consistent with a massive galaxy group. The group velocity is $\Delta v\,{=}\,{-}460{\pm}150$ \kms\ from the quasar and the velocity dispersion is $\sigma_{\rm group}\,{=}\,430{\pm}140$ \kms\ based on galaxies of $M_r\,{<}\,-20$ as shown in the inset panel Figure \ref{figure:HST} (with 2$\sigma$ clipping and uncertainties from bootstrap resampling). 
Not including the quasar, the light-weighted group center is ${\approx}8$ pkpc West and ${\approx}50$ pkpc South of the quasar.

\begin{sidewaysfigure*}
\vspace{-9cm}
\centering
	\includegraphics[width=\textwidth, angle=0]{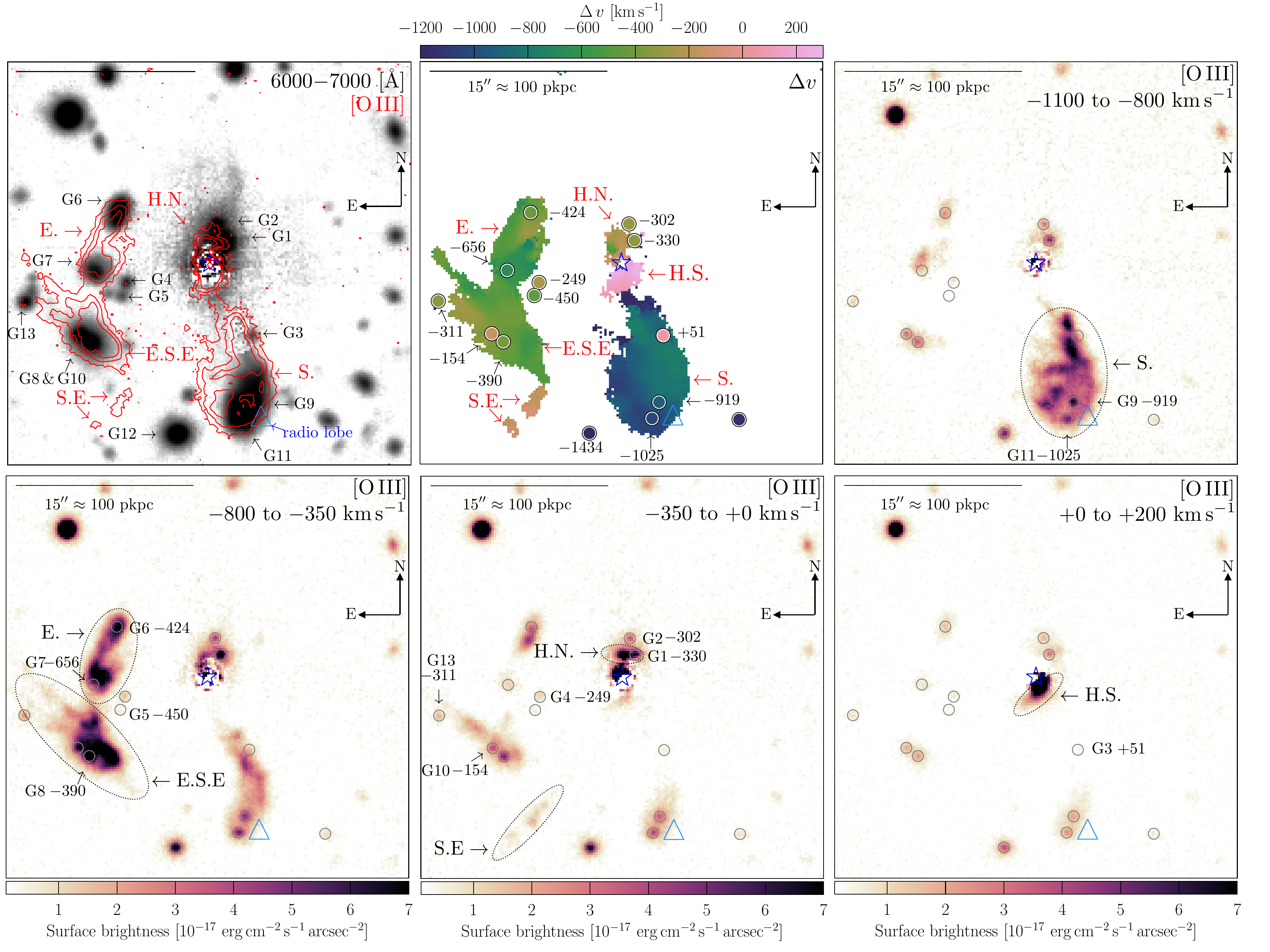}
	\caption{{\it Top left}: median flux over $6000{-}7000$ \AA\ in the MUSE datacube with [O\,III] surface brightness contours of $0.75$, 2.5, and $5$ ${\times}10^{-17}\,{\rm erg\,cm^{-2}\,s^{-1}\,{\rm arcsec}^{-2}}$ in red and with galaxies and nebulae labelled. {\it Top middle}: Map of the nebular line-of-sight velocities relative to the quasar. The positions of galaxies in the host group are marked by circles with line-of-sight velocity on the same scale. The {\it top right} panel and {\it bottom} row show narrow$-$band [O\,III] images extracted from the MUSE datacube over line-of-sight velocity intervals indicated in each panel. Nebulae are marked by ellipses and labelled while the quasar and radio lobe are marked by a blue star and triangle respectively. Galaxies are labelled if they fall within the displayed velocity interval.}
	\label{figure:narrows}
\end{sidewaysfigure*}

To gain insights into the environment of \pks, we display a $30''{\times}30''$ cutout of the quasar light subtracted MUSE image averaged over $6000{-}7000$ \AA\ (free of strong emission lines at $z\,{=}\,0.57$) in the top left panel of Figure \ref{figure:narrows}. The galaxy morphologies, projected separations, and relative velocities indicate that G6/G7, G9/G11, and G8/G10 are interacting galaxy pairs
with projected separations of $34$, $9$, and $7$ pkpc respectively.
G9/G11 are also nearly spatially coincident with one of the quasar radio lobes \citep[see][]{Sambruna:2004} which is labelled with a blue triangle in Figure \ref{figure:narrows}. All six of the interacting galaxies exhibit red rest-frame colors of $u\,{-}\,g\,{=}\,1.1$ to $1.7$.

G1/G2 are close projected pairs with one another ($10$ pkpc) and with the quasar (14 and 22 pkpc respectively). The quasar light subtracted MUSE image shown in the top left panel of Figure \ref{figure:narrows} reveals an arm of continuum emission extending ${\approx}60$ pkpc to the North of the quasar, a signature of recent or on-going interactions, possibly between the quasar host and G1/G2.

\begin{table*}
\caption{Summary of the properties of the ionized nebulae.}
\label{table:nebular}
\centering
\begin{tabular}{lrccrccccrc}
\hline
nebula       &  \multicolumn{1}{c}{ PA}  & major & minor &  \multicolumn{1}{c}{area} & \multicolumn{3}{c}{Total line luminosity} & \multicolumn{2}{c}{median} & associated\\ \cmidrule(lr){6-8} \cmidrule(lr){9-10}
                  &  (deg)                             & axis    & axis     & $(\rm pkpc^2)$                & [O\,III] & [O\,II] & H$\beta$ &  $\Delta v$ & \multicolumn{1}{c}{$\sigma$} & galaxies\\

                  &                                       & (pkpc) & (pkpc) &                                         & $({\rm erg\,s^{-1}})$ & $({\rm erg\,s^{-1}})$ & $({\rm erg\,s^{-1}})$ &  \multicolumn{1}{c}{$(\rm km\,s^{-1})$} & $(\rm km\,s^{-1})$ &  \\
\hline \hline
H.N.   & 355  & 24 & 10 & $ 190$ & $2.8{\times}10^{41}$ & $8.3\times10^{40}$ & $6.7{\times}10^{40}$ & ${-}230$ & $130$ &  host, G1, G2 \\
H.S.   &  40  & 34 & 10 & $ 280$ & $7.6{\times}10^{41}$ & $2.4\times10^{41}$ & $9.0{\times}10^{40}$ & ${+}150$ & $100$ &  host, G1, G2 \\
S.     &  92  & 72 & 48 & $2440$ & $3.4{\times}10^{42}$ & $1.3\times10^{42}$ & $5.6{\times}10^{41}$ & ${-}900$ & $ 70$ &  G9, G11      \\
S.E.   &  47  & 56 & 13 & $ 480$ & $9.2{\times}10^{40}$ & $2.6\times10^{40}$ & $8.8{\times}10^{39}$ & ${-}140$ & $ 50$ &  none          \\
E.S.E. & 317  & 96 & 33 & $2340$ & $2.1{\times}10^{42}$ & $1.0\times10^{41}$ & $2.4{\times}10^{41}$ & ${-}420$ & $ 90$ &  G8, G10      \\
E.     &  73  & 54 & 29 & $1430$ & $1.6{\times}10^{42}$ & $1.2\times10^{42}$ & $2.7{\times}10^{41}$ & ${-}500$ & $110$ &  G6, G7       \\
\hline
\end{tabular}
\end{table*}

\begin{figure}
	\centering
	\includegraphics[scale=0.68, angle=0]{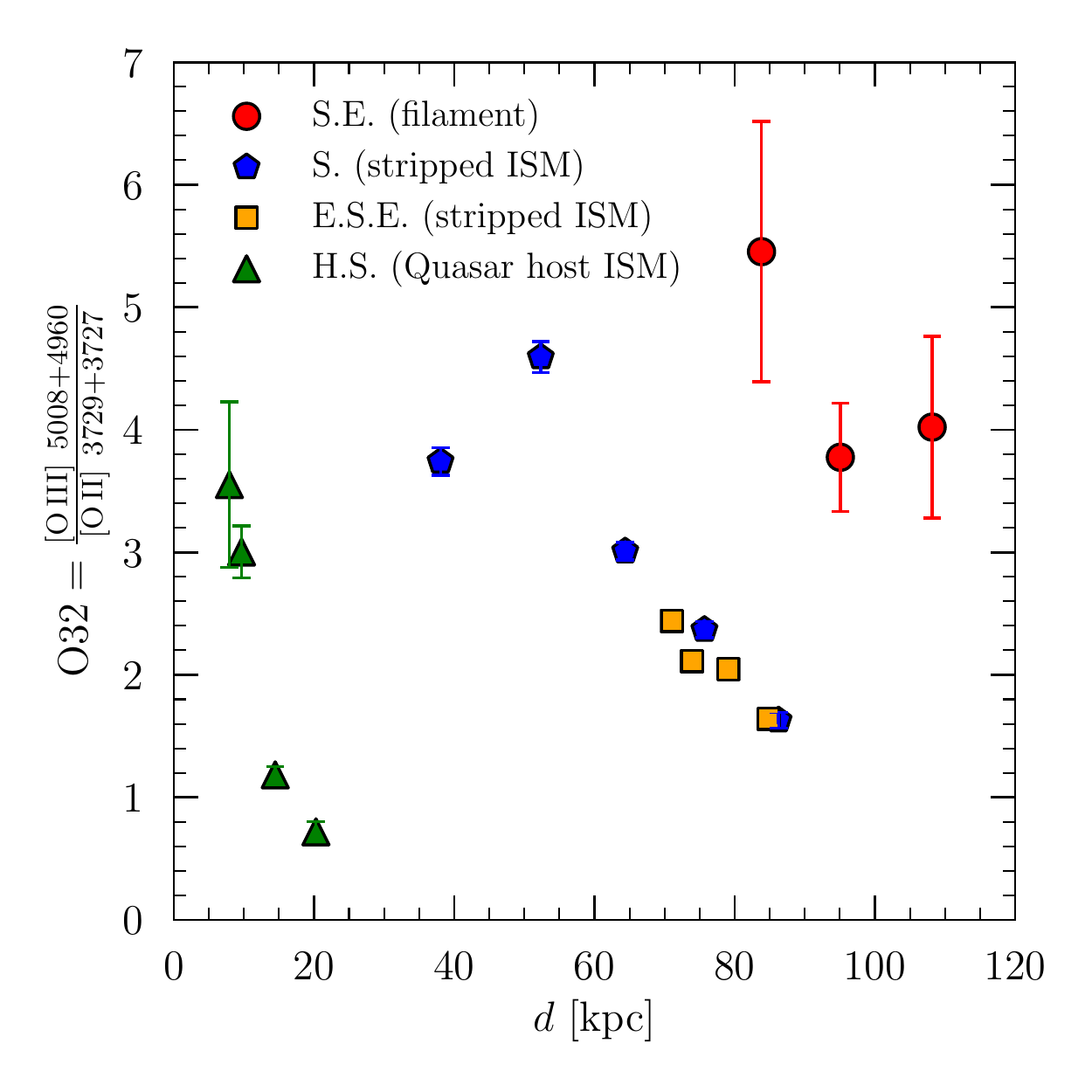}
	\caption{[O\,III] $5008{+}4960$ to [O\,II] $3729{+}3727$ emission line ratio as a function of projected distance from the quasar for the S.E., S., E.S.E., and H.S. nebulae as indicated in the figure legend. The H.N. and E. nebulae are not shown because they offer little leverage in projected distance.}
	\label{figure:O32}
\end{figure}

\section{Discovery and origins of $10{-}100$ pkpc scale ionized nebulae}
\label{section:nebulae}

The MUSE data enable the discovery of six ionized nebulae emitting strongly in [O\,III], [O\,II], and H$\beta$ on scales of $10{-}100$ pkpc and at line-of-sight velocities of $\Delta v\,{\approx}\,{-}1000$ to ${+}200$ \kms\ from the quasar. To visualize the morphologies of these nebulae along with galaxies in the group, we display [O\,III] emission contours over the MUSE wide-band image in the top left panel of Figure \ref{figure:narrows}. To visualize the kinematics of the nebulae and association with those of galaxies, the top middle panel of Figure \ref{figure:narrows} displays a velocity map of the nebulae from Gaussian fitting. Subsequent panels display narrow-band channels extracted from the datacube at the observed-frame wavelength of [O\,III] over velocities chosen to highlight each nebula and reveal detailed structure.

Here, we summarize the properties of the six nebulae and discuss their origins, proceeding from larger to smaller scales. Throughout, we refer to the nebulae by their position relative to the quasar as labelled in Figure \ref{figure:narrows}: South (S.), East-by-South East (E.S.E), East (E.), South East (S.E.), Host South (H.S.), and Host North (H.N.).
To quantify the properties of each nebula, we measured the line luminosity in [O\,III], [O\,II], and H$\beta$,  line-of-sight velocity relative to the quasar ($\Delta v$), and line-of-sight velocity dispersion ($\sigma$) by fitting Gaussian profiles to the emission lines at each spaxel with standard errors estimated from the MUSE error array and covariance matrix. We report the major axis position angle (PA); full extent along the major/minor axis; total line luminosity in [O\,III] ($5008{+}4960$), [O\,II] ($3727{+}3729$), and H$\beta$; median $\Delta v$; median $\sigma$; and associated galaxies in Table \ref{table:nebular}. Uncertainties in emission line luminosities, velocities, and velocity dispersions are ${<}15$\% and ${<}20$ \kms.

The nebulae exhibit high ionization states with mean $\rm [O\,III]/[O\,II]$ ratios of $1.3{-}3.5$ (Figure \ref{figure:O32}), [O\,III]/H$\beta$ of 4$-$10, and the brightest nebular regions exhibit He\,II $\lambda$4686 and [Ne\,V] $\lambda\lambda$3346, 3426 detections. Such high ionization states can be produced by photoionization by the quasar \citep[e.g.][]{Groves:2004} or fast shocks with velocities ${\gtrsim}400$ \kms\ \citep[e.g.][]{Allen:2008}. The median internal velocity dispersions of the nebulae are low ($50{-}130$ \kms) which disfavors the shock scenario. Moreover, [O\,III]/[O\,II] ratios within each nebula are generally anti-correlated with the projected distance from the quasar (Figure \ref{figure:O32}), consistent with gas in the quasar host group that is photoionized by the quasar. In future work (Johnson et al. in prep), we will present detailed studies of the physical conditions of the nebulae based on the full suite of available nebular diagnostics.

The three largest and most luminous nebulae (S., E.S.E., and E.) are kinematically and morphologically coincident with interacting galaxies in the field as shown in Figure \ref{figure:narrows}. In particular, S., E.S.E, and E. spatially and kinematically envelope interacting galaxy pairs G9/G11, G8/G10, and G6/G7 respectively.
The morphological and kinematic correspondence with interacting galaxies and the narrow internal velocity dispersions ($\sigma\,{=}\,70{-}110$ \kms) are most consistent with stripped ISM resulting from on-going interactions despite the spatial coincidence between the S. nebulae and a radio lobe \citep[c.f.][]{Harrison:2015}.

Unlike the other nebulae, S.E. is a narrow filament extending from $d\,{\approx}\,55$ to $120$ pkpc toward the quasar with a width of ${\approx}10$ pkpc (bottom middle panel of Figure \ref{figure:narrows}). S.E. is the faintest, lowest surface brightness, and most highly ionized of the nebulae despite being furthest from the quasar, suggesting that it has lower density.
S.E. exhibits a median velocity $\Delta v\,{=}\,{-}140$ \kms\ and a narrow median velocity dispersion of $50$ \kms. S.E. is not associated with any nearby galaxies detected in the HST or MUSE images which are complete for $m_{\rm F814W}{\lesssim}26.5$ mag sources ($M_r\,{=}\,{-}16$ at $z\,{=}\,0.57$). The morphology, calm kinematics, low surface brightness, high ionization state, and lack of any associated galaxies suggest that S.E. is a cool filament in the IGrM.

H.N. and H.S. are spatially and kinematically coincident with the quasar host at $d\,{\approx}\,10$ pkpc to the North and South and with median velocities of $\Delta v\,{=}\,{-}230$ and ${+}150$ \kms\ respectively. They extend over ${\approx}20{-}30$ pkpc and are morphologically reminiscent of tidal arms (bottom middle and right panels of Figure \ref{figure:narrows}). H.N. and H.S. and the extended arm of continuum emission seen North of the host may be the result of on-going interactions with G1/G2 or another more advanced merger. The H.N. and H.S. nebulae exhibit similar [O\,III] to [O\,II] ratios to the other nebulae despite being significantly closer to the quasar. This is consistent with expectations for the ISM of the quasar host which is likely to have higher densities than either stripped ISM or cool filaments. We do not detect any nebular emission associated with the continuum arm. Adopting the measured velocities and projected distances of H.N. and H.S., we estimate a dynamical mass for the quasar host of $M_{\rm dyn}({<}10\ {\rm pkpc})\,{\sim}\,{10^{11}\,{\rm M}_\odot}/{\sin^2 i}$ where $i$ is the inclination angle (assuming $v^2(r)\,{\approx}\,GM(<r)/r$).

\section{Discussion and conclusions}
\label{section:discussion}

With deep MUSE observations, we discovered six ionized nebulae on scales of $10{-}100$ pkpc in the $z\,{\approx}\,0.57$ galaxy group hosting \pks, one of the most luminous quasars at $z\,{<}\,1$. Although the nebulae are distributed over a wide range of position angles and projected distances from the quasar, all exhibit line ratios that are most consistent with quasar photoionization, requiring  a large opening angle and an active lifetime of ${\gtrsim}3{\times}10^5$ yr (the light travel time for ${\approx}100$ kpc). 
The quasar photoionization enables observations of cool IGrM and ISM that would otherwise be neutral and not observable in emission beyond the local Universe. When combined with the morpho-kinematic information from a wide-field IFS, this enables new insights into galaxy and quasar fuel supplies from the IGrM to the ISM.

One of the nebulae (S.E.) exhibits a filamentary morphology extending $50$ pkpc toward the quasar with narrow internal velocity dispersion ($50$ \kms) and is not associated with any detected galaxies. These properties are most consistent with a cool filament in the IGrM like those observed more locally in groups \citep[e.g.][]{Xu:1999} and in cool-core clusters \citep[e.g.][]{McDonald:2010, Gaspari:2018}. Such filaments often occur in central networks but are also observed in what would appear as single filaments in our seeing limited data (e.g. Abell 1796). The truncation of the filamentary nebulae before reaching the quasar host would be unusual in a cluster but may result from increased ionization state at smaller distance from the quasar weakening optical emission lines.

The superb imaging and spectral quality of MUSE enable subtraction of the quasar light to reveal a 60 kpc arm of continuum emission  and two ${\approx}10$ pkpc tidal-arm like nebulae extending from the quasar host, all signatures of interactions.
The presence of a cool filament extending toward the host, continuum arm, and nebulae near the host make \pks\ one of the strongest candidates for an interaction triggered quasar and an ideal case-study of the relationship between galaxy/black hole growth and gas supplies on scales from the IGrM (${\approx}100$ pkpc) to the ISM (${\approx}10$ pkpc). 
In particular, the filamentary nebula (S.E.) may have supplied the ISM of the quasar host with cooling IGrM while interactions may redistribute angular momentum in ISM to fuel nuclear activity.

The three largest nebulae (S., E.S.E, and E.) are among the largest and most luminous [O\,III] ``blobs'' known \citep[e.g.][]{Schirmer:2016, Yuma:2017, Epinat:2018}. They exhibit large line-of-sight velocities relative to the quasar (${-}900$ to ${-}400$ \kms) but narrow internal velocity dispersions ($70{-}110$ \kms). Joint morphological and kinematic analyses indicate that these three  nebulae originate from ISM stripped from interacting galaxies in the host group rather than from outflows as often assumed \citep[e.g.][]{Fu:2009, Yuma:2017}. The stripping may be due to tidal forces \citep[e.g.][]{Marasco:2016} or ram pressure \citep[e.g.][]{Hester:2006} which can remove H\,I disks and decrease star formation in even massive satellites that interact with one another or pass through central, dense IGrM regions. The velocity shear, one-sided morphology and sharp truncation of the S. nebulae are suggestive of ram pressure stripping though variability or partial obscuration of the quasar may also be responsible. Such large nebulae from stripped material could potentially explain the densest components of extended nebulae observed around luminous quasars at $z\,{>}\,2$ \citep[e.g.][]{Borisova:2016, Law:2018}.

The discovery of extended, ionized nebulae including a cool filament and interaction related debris in the environment of \pks\ demonstrate the power of wide-field IFS for studying quasar hosts. The nebulae cover 20\% of the area around the quasar at $\lesssim100$ kpc
which may explain the high covering fraction of absorbing gas around luminous quasar hosts \citep[][]{Johnson:2015b} if the gas extends to larger distances at lower densities. In future work (Johnson et al. in preparation), we will present a survey of the ionized nebulae around quasar hosts from the full MUSE-QuBES sample including the full suite of available nebular diagnostic.

\section*{Acknowledgements}
We are grateful to the anonymous referee, J. Greene, M. Gaspari, F. Schweizer, and M. Strauss for suggestions that improved paper.
SDJ is supported by a NASA Hubble Fellowship (HST-HF2-51375.001-A).
SC acknowledges support from Swiss National Science Foundation grant PP00P2\_163824.

Based on observations from the European Organization for Astronomical Research in the Southern Hemisphere under ESO (PI: Schaye, PID: 094.A-0131) and the NASA/ESA Hubble Space Telescope (PI: Mulchaey, PID: 13024).
The paper made use of the NASA/IPAC Extragalactic Database, the NASA Astrophysics Data System, Astropy \citep[][]{The-Astropy-Collaboration:2018}, and Aplpy \citep[][]{Robitaille:2012}.

\end{document}